\documentclass[9pt]{article}
\usepackage{spconf}
\usepackage[utf8]{inputenc}
\usepackage{amsmath,amsthm,amssymb,algorithm,tikz}
\usepackage{algpseudocode}
\usepackage{enumitem,epsfig}
\usepackage{adjustbox}
\usepackage{graphicx}
\usepackage{rotating}
\usepackage{multirow}
\usepackage{subcaption}
\usepackage{multicol}
\usepackage{float}
\usepackage{hyperref}
\usepackage{wrapfig}
\usepackage{bm}

\usetikzlibrary{positioning}

\DeclareMathOperator*{\argmin}{argmin}

\DeclareRobustCommand*\textsubscript[1]{%
  \@textsubscript{\selectfont#1}}
\def\@textsubscript#1{%
  {\m@th\ensuremath{_{\mbox{\fontsize\sf@size\z@#1}}}}}

\def\tim{\times_3^1}

\def\L{\mathbf{L}}
\def\R{\mathbf{R}}
\def\A{\mathcal{A}}
\def\B{\mathcal{B}}
\def\C{\mathcal{C}}

\def\N{\mathcal{N}}
\def\T{\mathbf{T}}

\def\U{\mathcal{U}}
\def\V{\mathcal{V}}
\def\Z{\mathcal{W}}
\def\S{\mathcal{S}}

\def\P{\bm{\pi}}
\def\X{\mathcal{X}}
\def\Y{\mathcal{Y}}
\def\Z{\mathcal{Z}}

\def\tdst{\times_3^1\dots\times_3^1}


\title{Graph Regularized Tensor Train Decomposition}
\name{Seyyid Emre Sofuoglu and Selin Aviyente \thanks{This work was in part supported by NSF CCF-1615489.}}
\address{Department of Electrical and Computer Engineering, Michigan State University, East Lansing, MI.}

\begin{document}
\makeatletter
\let\origsection\section
\let\origsubsection\subsection
\let\origsubsubsection\subsubsection
\renewcommand\section{\@ifstar{\starsection}{\nostarsection}}
\renewcommand\subsection{\@ifstar{\starsubsection}{\nostarsubsection}}
\renewcommand\subsubsection{\@ifstar{\starsubsubsection}{\nostarsubsubsection}}

\newcommand\nostarsection[1]
{\sectionprelude\origsection{#1}\sectionpostlude}

\newcommand\starsection[1]
{\sectionprelude\origsection*{#1}\sectionpostlude}

\newcommand\nostarsubsection[1]
{\sectionprelude\origsubsection{#1}\sectionpostlude}

\newcommand\starsubsection[1]
{\sectionprelude\origsubsection*{#1}\sectionpostlude}

\newcommand\nostarsubsubsection[1]
{\sectionprelude\origsubsubsection{#1}\sectionpostlude}

\newcommand\starsubsubsection[1]
{\sectionprelude\origsubsubsection*{#1}\sectionpostlude}

\newcommand\sectionprelude{%
  \vspace{-1.3em}
}

\newcommand\sectionpostlude{%
  \vspace{-.8em}
}
\makeatother

\ninept
\maketitle
\begin{abstract}
    With the advances in data acquisition technology, tensor objects are collected in a variety of applications including multimedia, medical and hyperspectral imaging. As the dimensionality of tensor objects is usually very high, dimensionality reduction is an important problem. Most of the current tensor dimensionality reduction methods rely on finding low-rank linear representations using different generative models. However, it is well-known that high-dimensional data often reside in a low-dimensional manifold. Therefore, it is important to find a compact representation, which uncovers the low-dimensional tensor structure while respecting the intrinsic geometry. In this paper, we propose a graph regularized tensor train (GRTT) decomposition that learns a low-rank tensor train model that preserves the local relationships between tensor samples. The proposed method is formulated as a nonconvex optimization problem on the Stiefel manifold and an efficient algorithm  is proposed to solve it. The proposed method is compared to existing tensor based dimensionality reduction methods as well as tensor manifold embedding methods for unsupervised learning applications.
\end{abstract}
\section{Introduction}
In a lot of emerging applications ranging from computer vision to hyperspectral imaging, data are captured as higher order structures. Traditional unsupervised and supervised learning methods developed for 1-D signals do not translate well to higher order data structures as they get computationally prohibitive. For this reason, recent years have seen a growth in the development of tensor decomposition methods for dimensionality reduction. These methods include extensions of PCA, SVD and NMF to tensors including PARAFAC/CP, Tucker decomposition and Tucker-NMF \cite{sidiropoulos2017tensor, cichocki2014era, kolda2009tensor, bro1997parafac,de2000multilinear,de2000best,cichocki2015tensor,lee2016nonnegative,wang2018principal}. However, these methods generally fail to consider the geometric  structure of data. The geometric relationship between data samples has been shown to be important for learning low-dimensional structures from high-dimensional data \cite{tenenbaum2000global,wang2018tensor}.

Recently, motivated by manifold learning, dimensionality reduction of tensor objects has been formulated to incorporate the geometric structure \cite{li2016mr}. The goal is to learn a low dimensional representation for tensor objects that incorporates the geometric structure  while maintaining a low reconstruction error in the tensor decomposition. This idea of manifold learning for tensors has been mostly implemented for the Tucker method, including Graph Laplacian Tucker Decomposition (GLTD) \cite{jiang2018image} and nonnegative Tucker factorization (NTF). However, this line of work suffers from the limitations of Tucker decomposition such as the exponential increase in the storage cost \cite{cichocki2017tensor}.

Tensor-Train (TT) model, on the other hand, provides better compression than Tucker models, especially for higher order tensors, as it expresses a given high-dimensional tensor as the product of low-rank, 3-mode tensors \cite{cichocki2014era}. TT model has been employed in various applications such as PCA \cite{bengua2017matrix}, manifold learning \cite{wang2018tensor} and deep learning \cite{novikov2015tensorizing}. In our previous work, we proposed a TT model such that the features can be matrices and higher-order tensors, and this way, the computational efficiency could be improved \cite{sofuoglu2019multi}. In this paper, we propose a graph-regularized TT decomposition for unsupervised dimensionality reduction which utilizes the two-way approach. 

This paper differs from the current work in three ways. First, existing TT methods focus solely on obtaining low-rank approximations that minimize the reconstruction error without preserving the intrinsic geometry or on obtaining preserving the intrinsic geometry without reconstruction error in mind. Second, existing tensor graph regularized methods are mostly focused on Tucker based models which are known to be inefficient in terms of storage cost and computational complexity. Finally, the current paper proposes a novel ADMM based algorithm to solve the nonconvex optimization method which makes it much faster than existing graph regularized tensor decomposition methods.

\section{Background}

\label{sec:back}
\noindent  For a given dataset $\Y \in \mathbb{R}^{I_1 \times \dots\times I_N \times S}$ with $S$ samples, let $\Y_s\in \mathbb{R}^{I_1 \times I_2 \times\dots\times I_N }$ be a tensor where $s\in\{1,\dots,S\}$ is the index.

\subsection{Tensor Operations}
\noindent \textbf{Definition 1.} (Reshaping) $\mathbf{T}_n(.)$ is a tensor-to-matrix reshaping operator defined as $\mathbf{T}_n(\Y_s)\in \mathbb{R}^{I_1\dots I_n\times I_{n+1}\dots I_N}$.

\noindent \textbf{Definition 2.} (Left and right unfolding) The left unfolding operator creates a matrix from a tensor by taking all modes except the last mode as row indices and the last mode as column indices, i.e. $\mathbf{L}(\Y_s) \in \mathbb{R}^{I_1I_2\dots I_{N-1}\times I_N}$ which is equivalent to $\T_{N-1}(\Y_s)$. Right unfolding transforms a tensor to a matrix by taking all the first mode fibers as column vectors, i.e. $\mathbf{R}(\Y_s) \in \mathbb{R}^{I_1\times I_2I_3\dots I_N}$ which is equivalent to $\T_1(\Y_s)$. A tensor is defined to be left (right)-orthogonal if its left (right) unfolding is orthogonal.

\noindent \textbf{Definition 3.} (Tensor Merging Product) Tensor merging product connects two tensors along some given sets of modes. For two tensors $\A\in \mathbb{R}^{I_1\times I_2\times\dots\times I_N}$ and $\B\in \mathbb{R}^{J_1\times J_2\times\dots\times J_M}$ where $I_n=J_m$ and $I_{n+1}=J_{m-1}$ for some $n$ and $m$, tensor merging product \cite{cichocki2017tensor} is given as
$\C=\A\times_{n,n+1}^{m,m-1}\B$, where $\C\in\mathbb{R}^{I_1\times\dots \times I_{n-1}\times I_{n+2}\times \dots \times I_N\times J_1\times\dots\times J_{m-2}\times J_{m+1}\times\dots\times J_M}$ is calculated as:

\vspace{-1em}
\small
\begin{gather}
\C(i_1,\dots , i_{n-1}, i_{n+2}, \dots , i_N, j_1,\dots, j_{m-2}, j_{m+1},\dots, j_M)= \nonumber \\ 
\sum_{t_1=1}^{I_n}\sum_{t_2=1}^{J_{m-1}}\big[\A(i_1,\dots,i_{n-1},i_n=t_1,i_{n+1}=t_2,i_{n+1},\dots,i_N)\nonumber \\ \B(j_1,\dots,j_{m-2},j_{m-1}=t_2,j_m=t_1,j_{m+1},\dots,j_M)\big].
\end{gather}
\normalsize

\subsection{Tensor-Train Decomposition} 
\noindent Using MPS, each element of $\Y_s$ can be represented as \cite{bengua2017matrix}:
\begin{gather}
\Y_s(i_1,i_2,\dots,i_N)=\U_1(1,i_1,:)\U_2(:,i_2,:) \dots \U_k(:,i_k,:)\nonumber \\X_s\U_{k+1}(:,i_{k+1},:) \dots \U_N(:,i_N,:),
\label{eq:mps}
\end{gather}
where $\U_n\in \mathbb{R}^{r_{n-1}\times I_n \times r_n}$ for $n\leq k$, $\U_n\in \mathbb{R}^{r_{n}\times I_n \times r_{n+1}}$ for $n>k$ are the tensor factors. $r_n<I_n, \forall n \in\{1,\dots,N\}$ and correspond to the ranks of the different modes. $X_s\in \mathbb{R}^{r_k\times r_{k+1}}$ is the low dimensional projection matrix. The parameter $k$ is selected using a center of mass approach to reduce computational complexity, \textit{i.e.} $k$ is selected such that $|\prod_{i=1}^k I_i-\prod_{j=k+1}^N I_{j}|$ is minimized. TT decomposition given in (\ref{eq:mps}) can be rewritten as, $\Y_s=\U_1\tdst\U_k\times_3^1X_s\times_2^1\U_{k+1}\tdst\U_N$ \cite{cichocki2017tensor}.

When $\Y_s$ is reshaped into a matrix, (\ref{eq:mps}) can be equivalently expressed as a matrix projection $\T_k(\Y_s)=\left[\mathbb{I}_{I_n} \otimes U_{\leq k} \right]X_s U_{>k}$,
where $U_{\leq k}=\L(\U_1\times_3^1\U_2\times_3^1\dots \times_3^1\U_k)\in \mathbb{R}^{I_1I_2\dots I_k\times r_k}$ and $U_{>k}=\R(\U_{k+1}\tdst\U_N)\in \mathbb{R}^{r_{k+1} \times I_{k+1}\dots I_N}$. When $\L(\U_n)$s for $n\leq k$ are left orthogonal, $U_{\leq k}$ is also left orthogonal \cite{holtz2012manifolds}. Similarly, when $\R(\U_n)$s for $n>k$ are right orthogonal, $U_{>k}$ is right orthogonal.



\section{Graph Regularized Tensor-Train}
Our goal is to find a TT projection such that the geometric structure of the samples $\Y_s$ is preserved, i.e. the distance between the samples, $\Y_s$, should be similar to that between the projections $X_s$, while the reconstruction error of the low-rank TT decomposition is minimized.  This goal can be formulated through the following cost function as:
\vspace{-.5em}
\begin{gather}
    f_O(\{\U\},\X)=\nonumber\\\sum_{s=1}^{S}\|\Y_s-\U_1\tdst\U_k\tim X_s\times_2^1\U_{k+1}\tdst\U_N\|_F^2\nonumber \\+\frac{\lambda}{2}\sum_{s=1}^S\sum_{\substack{s'=1\\ s'\neq s}}^S\|X_s-X_{s'}\|_F^2 w_{ss'},  \quad \L(\U_n)^\top\L(\U_n)=\mathbb{I}_{r_n} , \forall n \nonumber
\end{gather}
where $\{\U\}$ denotes the set of tensor factors $\U_n, \forall n\in \{1,\dots,N\}$, $\X\in \mathbb{R}^{r_k\times S\times r_{k+1}}$ is the tensor whose slices are $X_s$ and $w_{ss'}$ is the similarity between tensor samples defined by:
\vspace{-.5em}
\begin{gather}
    w_{ss'}=
    \begin{cases}
        1, & \text{if} \quad \Y_s\in \N_k(\Y_{s'})\quad \text{or}\quad \Y_{s'}\in \N_k(\Y_{s})\\
        0, & \text{otherwise}
    \end{cases},
\end{gather}
where $\N_k(\Y_s)$ is the k-nearest neighborhood of $\Y_s$.

The objective function can equivalently be expressed as:
\begin{gather}
    f_O(\{\U\},\X)=\nonumber\\ \|\P_{k+1}(\Y)-\U_1\tdst\U_k\tim \X\tim\U_{k+1}\tdst\U_N\|_F^2\nonumber \\+\lambda \mathrm{tr}\left((\X\times_2^1 L)\times_{1,3}^{1,2}\X\right), \nonumber \\
    \L(\U_n)^\top\L(\U_n)=\mathbb{I}_{r_n}, \text{ for } n\leq k \nonumber\\ \text{ and } \R(\U_n)\R(\U_n)^\top=\mathbb{I}_{r_{n}}, \text{ for } n> k,\nonumber
\end{gather}
where $\P_{k+1}(\Y)\in\mathbb{R}^{I_1\times\dots\times I_k\times S\times I_{k+1}\times\dots\times I_N}$ is the permuted version of  $\Y$ such that the last mode is moved to the $(k+1)$th mode and all  modes larger than $k$ are shifted by one mode,  $W\in \mathbb{R}^{S\times S}$ is the adjacency matrix and $L=D-W \in \mathbb{R}^{S\times S}$ is the graph Laplacian where $D$ is a diagonal degree matrix with,  $d_{ss}=\sum_{s'=1}^S w_{ss'}$.

\subsection{Optimization}

The goal of obtaining low-rank tensor train projections that preserve the data geometry can be achieved by minimizing the objective function as follows: 
\vspace{-.5em}
\begin{gather}
    \argmin_{\{\U\},\X}f_{O}(\{\U\},\X), \text{ s.t. } \L(\U_n)^\top\L(\U_n)=\mathbb{I}_{r_n}, \text{ for } n\leq k,\\ \text{ and } \R(\U_n)\R(\U_n)^\top=\mathbb{I}_{r_{n}}, \text{ for } n> k.\nonumber
\end{gather}
As we want our tensor factors to be orthogonal, the solutions lie in the Stiefel manifold $\S_n$ , i.e. $\L(\U_n) \in \S_n$ for $n\leq k$ and $\R(\U_n)^\top \in \S_n$ for $n>k$. Although the function $f_O(.)$ is convex, the optimization problem is nonconvex due to the manifold constraints on $\U_n$s.

The solution to the optimization problem can be obtained by Alternating Direction Method of Multipliers (ADMM). In  order  to  solve  the  optimization  problem we define $\{\V\}$, as the set of auxiliary variables $\V_n, \forall n\in\{1,\dots,N\}$ and rewrite the objective function as:
\vspace{-.5em}
\begin{gather}
    \argmin_{\{\U\},\{\V\},\X}f_{O}(\{\V\},\X) \qquad \text{subject to} \quad \U_n=\V_n \quad, \forall n \nonumber \\ \L(\U_n) \in \S_n,  \forall n\leq k \text{ and } \R(\U_n)^\top \in \S_n, \forall n>k.\nonumber
\end{gather}
The partial augmented Lagrangian is given by:
\vspace{-.5em}
\begin{gather}
    \mathcal{L}\left(\{\U\},\{\V\},\X,\{\Z\}\right) = f_{O}(\{\V\},\X) -\nonumber \\\sum_{n=1}^N\Z_n\times_{1,2,3}^{1,2,3}(\V_n-\U_n)+\frac{\gamma}{2}\sum_{n=1}^N\|\V_n-\U_n\|_F^2,
\end{gather}
where $\Z_n$s are the Lagrange multipliers and $\gamma$ is the penalty parameter.

As each tensor factor is independent from the others, we update the variables for each mode $n$ using the corresponding part of the augmented Lagrangian:
\vspace{-.5em}
\begin{gather}
    \mathcal{L}_n\left(\U_n,\V_n,\Z_n \right)= f_O(\V_n)-\nonumber \\\Z_n\times_{1,2,3}^{1,2,3}(\V_n-\U_n)+\frac{\gamma}{2}\|\V_n-\U_n\|_F^2,
    \label{eq:partauglag}
\end{gather}
where $f_O(\V_n)$ denotes the objective function where all variables other than $\V_n$ are fixed. The solution for each variable at iteration $t+1$ can then be found using a step-by-step approach as:
\vspace{-.5em}
\begin{gather}
    \V_{n}^{t+1}=\argmin_{\V_n} \mathcal{L}_n\left(\U_n^t,\V_n,\Z_n^{t}\right),\\
    \U_n^{t+1}=\argmin_{\U_n}\begin{cases}
    \mathcal{L}_n\left(\U_n,\V_n^{t+1},\Z_n^t\right), \L(U_n)\in \S_n &\text{ for } n\leq k, \\
    \mathcal{L}_n\left(\U_n,\V_n^{t+1},\Z_n^t\right), \R(U_n)^\top\in \S_n &\text{ for } n> k,
    \end{cases}
    \nonumber \\
    \Z_n^{t+1}=\Z_n^t-\gamma(\V_n^{t+1}-\U_n^{t+1}).
    \label{eq:optZ}
\end{gather}
Once $\V_n, \U_n, \Z_n$ are updated for all $n$, samples $\X$ are computed using:
\vspace{-.5em}
\begin{gather}
    \X^{t+1} = \argmin_{\X} \mathcal{L}\left(\{\U^{t+1}\},\{\V^{t+1}\},\X,\{\Z^{t+1}\}\right).
    \label{eq:optX}
\end{gather}

\subsubsection{Solution for $\V_n$}
For $n\leq k$, the solution for $\V_n^{t+1}$ can be written explicitly as:
\vspace{-.5em}
\begin{gather}
    \V_n^{t+1}=\argmin_{\V_n} \|\P_{k+1}(\Y)-\nonumber\\\V_1^{t+1}\tdst \V_n\tdst\V_k^t \tim\X^t\tim\V_{k+1}^t\tdst\V_N^t\|_F^2\nonumber \\ -\Z_n^t\times_{1,2,3}^{1,2,3}(\V_n-\U_n^t)+\frac{\gamma}{2}\|\V_n-\U_n^t\|_F^2.
\end{gather}
We can equivalently convert this equation into matrix form as:
\vspace{-.5em}
\begin{gather}
    \V_n^{t+1}= \argmin_{\V_n} \left\|H \L(\V_n)P -\mathbf{T}_n(\P_{k+1}(\Y))\right\|_F^2\nonumber \\- \mathrm{tr}\Big(\L(\Z_n^t)^\top\L(\V_n-\U_n^t)\Big) + \frac{\gamma}{2}\|\L(\V_n)-\L(\U_n^t)\|_F^2,
    \label{eq:OptMin}
\end{gather}
where $H=\left[\mathbb{I}_{I_n} \otimes V_{\leq n-1}^{t+1} \right]$, $P=\R(\V_{n+1}^t\tdst\V_k^t\tim\X^t\tdst\V_N^t)$. The analytical solution is found by taking the derivative with respect to $\L(\V_n)$ and setting it to zero:
\begin{gather}
    2H^\top\Bigg(H\L(\V_n^{t+1})P-G\Bigg)P^\top-\L(\Z_n^t) + \gamma\L(\V_n^{t+1})-\gamma\L(\U_n^t)=0, \nonumber \\
    \T_3(\V_n^{t+1})=\left(2(PP^\top\otimes H^\top H)+\gamma\mathbb{I}_{r_{n-1}I_nr_n}\right)^{-1}\nonumber \\ \left(\T_3(\gamma\U_n^t+\Z_n^t)+2\T_2(H^\top GP^\top)\right),
    \label{eq:unsolnl}
\end{gather}
where $G=\mathbf{T}_n(\P_{k+1}(\Y))$. Note that the inverse in the solution will always exist given $\gamma>0$ as the inverse of a sum of a Hermitian matrix and an identity matrix always exists.

When $n>k$, following (\ref{eq:OptMin}) the solution for $\V_n$ can be written in the same manner but with different $H, G$ and $P$,
where $H=V_{\leq n-1}^{t}$, $P=\left[V_{>n}^{t+1} \otimes \mathbb{I}_{I_n}\right]$ and $G=\mathbf{T}_{n+1}(\P_{k+1}(\Y))$.

\subsubsection{Solution for $\U_n$}
For $n\leq k$, we can solve (\ref{eq:partauglag}) for $\U_n$ using:
\begin{gather}
    \U_n^{t+1}=\argmin_{\U_n:\L(\U_n)\in \S_n}-\mathrm{tr}\Big(\L(\Z_n^t)^\top\L(\V_n^{t+1}-\U_n)\Big) \nonumber \\+ \frac{\gamma}{2}\|\L(\V_n^{t+1})-\L(\U_n)\|_F^2=\nonumber \\ \argmin_{\U_n}\|\L(\V_n^{t+1})-\frac{1}{\gamma}\L(\Z_n^t)-\L(\U_n)\|_F^2,
    \label{eq:vnsolnl}
\end{gather}
which is found by applying a singular value decomposition to $\L(\V_n^{t+1})-\frac{1}{\gamma}\L(\Z_n^t)$. When $n>k$, the optimal solution is similarly found by applying SVD to $\R(\V_n^{t+1})-\frac{1}{\gamma}\R(\Z_n^t)$.

\subsubsection{Solution for $\X$}
Let $\P_2(\X)\in \mathbb{R}^{r_k\times r_{k+1}\times S}$ be the permutation of $\X$, (\ref{eq:optX}) can equivalently be rewritten in matrix form as:
\begin{gather}
    \argmin_{\X}\left\|\left[ {V_{>k}^{t+1}}^\top \otimes V_{\leq k}^{t+1} \right] \L(\P_2(\X)) -\mathbf{T}_N(\Y)\right\|_F^2 \nonumber \\ +\lambda \mathrm{tr}\left(\L(\P_2(\X)) L\L(\P_2(\X))^\top\right).
    \label{eq:optXM}
\end{gather}
The solution for $\X^{t+1}$ does not have any constraints, thus it is solved analytically by setting the derivative of ($\ref{eq:optXM}$) to zero:
\begin{gather}
    2H^\top(H\L(\P_2(\X^{t+1}))-G)+2\lambda\L(\P_2(\X^{t+1}))L=0, \nonumber \\
    H^\top H\L(\P_2(\X^{t+1}))+\lambda\L(\P_2(\X^{t+1}))L = H^\top G, 
    \label{eq:upX}
\end{gather}
where $H={V_{>k}^{t}}^\top \otimes V_{\leq k}^{t+1}$ and $G=\mathbf{T}_N(\Y)$. (\ref{eq:upX}) is a Sylvester equation which can be solved efficiently \cite{bartels1972solution}. Similar to the case for $\V_n$, the solution to this problem always exists.
\subsubsection{Solution for $\Z_n$}
Finally, we update the Lagrange multipliers $\Z_n^t$ using (\ref{eq:optZ}).

\algrenewcommand\algorithmicrequire{\textbf{Input:}}
\algrenewcommand\algorithmicensure{\textbf{Output:}}
\begin{algorithm}
\caption{Graph Regularized Tensor Train-ADMM(GRTT-ADMM)}
\begin{algorithmic}[1]
\Require Input tensors $\Y_s \in \mathbb{R}^{I_1 \times I_2 \times \dots \times I_N }$ where $s \in \{1,\dots ,S\}$, initial tensor factors $\{\U^1\},  n \in \{1,\dots ,N\}$, $k$, $\lambda$, $r_1,\dots,r_N$, $LoopIter$, $ConvThresh$
\Ensure $\U_n, n \in \{1,\dots ,N\}$, and $X_s,\quad \forall s$
\State $\{\V^1\} \gets \{\U^1\}$.
\State $\{\Z^1\} \gets 0$.
\While{ $t<LoopIter$ \textbf{or} $c>ConvThres$}
\For{$n=1$ : $N$}
\State Find $\V_n^{t+1}$ using (\ref{eq:unsolnl}).
\State Find $\U_n^{t+1}$ using SVD to solve (\ref{eq:vnsolnl}).
\State Find $\Z_n^{t+1}$ using (\ref{eq:optZ}).
\EndFor
\State Find $\X^{t+1}$ using (\ref{eq:upX}).
\State $c \gets \frac{1}{N}\sum_{n=1}^N\frac{\|\V_n^{t+1}-\V_n^{t}\|_F^2}{\|\V_n^t\|_F^2}$
\State $t=t+1$.
\EndWhile
\end{algorithmic}
\end{algorithm}

\subsection{Convergence}
Convergence of ADMM is guaranteed for convex functions but there is no theoretical proof of the convergence of ADMM  for nonconvex functions. Recent research has provided some theoretical guarantees for the  convergence of ADMM for a class of nonconvex problems  under some conditions \cite{wang2019global}. 


Our objective function is nonconvex due to unitary constraints. In \cite{wang2019global}, it has been shown that this type of nonconvex optimization problems, i.e. convex optimization on a Stiefel manifold, converge under some conditions. We show that these conditions hold for each optimization problem corresponding to mode $n$. The gradient of $f_O$ with respect to $\V_n$ is Lipschitz continuous with $L\geq\|P P^\top \otimes H^\top H \|_2$, which fulfills the conditions given in \cite{wang2019global}. Thus, $\mathcal{L}_n$ convergences to a set of solutions $\V_n^t, \U_n^t, \Z_n^t,$ given that $\gamma\geq 2L+1$. The solution for $\X$ is  found analytically. As the iterative solutions for each variable converge and the optimization function is nonnegative, \textit{i.e.} bounded from below, the algorithm converges to a local minimum.

\section{Experiments}
The proposed method is evaluated for clustering and compared to existing tensor clustering methods including k-means, MPS \cite{bengua2017matrix}, TTNPE \cite{wang2018tensor} and GLTD \cite{jiang2018image} for Weizmann Face Database and MNIST Dataset. Clustering quality is quantified by Normalized Mutual Information (NMI). Average accuracy with respect to both storage complexity and computation time over 20 experiments are reported for all methods. 


In the following experiments, the storage complexity is quantified as the size of the tensor factors ($\U_{n}, \forall n$) and projections ($\mathcal{X}_{s}, \forall s$). The varying levels of storage cost are obtained by varying $r_{n}$s in the implementation of the tensor decomposition methods. Using varying levels of a truncation parameter $\tau \in (0,1]$, the singular values smaller than $\tau$ times the largest singular value are discarded. The rest are used to determine ranks $r_n$ for both TT-based and TD-based methods. For GRTT and TTNPE, the ranks are selected using TT-decomposition proposed in \cite{oseledets2011tensor}, while for GLTD truncated HOSVD was used. Computational complexity is quantified as the time it takes to learn the tensor factors.  In order to compute the run time, for TT-based methods, each set of tensor factors is optimized until the change in the normalized difference between consecutive tensor factors is less than $0.01$ or $50$ iterations are completed.

The regularization parameter, $\lambda$, for each experiment was selected using a validation set composed of a small batch of samples not included in the experiments. 5 random experiments were conducted and optimal $\lambda$ was selected as the value that gave the best average NMI for a range of $\lambda$ values from $0.001$ to $1000$ increasing in a logarithmic scale. The similarity graphs were constructed using k-nearest neighbor method with $k=log(S)$ following \cite{von2007tutorial}. 


\subsection{MNIST}
MNIST is a database of grayscale handwritten digit images where each image is of size $28\times 28$. We transformed each of the images to a $4\times 7\times4\times 7$ tensor. Reshaping the inputs into higher order tensors is common practice and was employed in prior work \cite{khoromskij2011dlog, oseledets2011tensor, zhao2016tensor, cichocki2017tensor, bengua2017efficient}. In our experiments, we used a subset of 500 images with 50 images from each class. 50 samples with 5 samples from each class are used as validation set to determine $\lambda$.

\begin{figure}[htb]
\centering
\begin{subfigure}[b]{.9\columnwidth}
    \centering
    \includegraphics[width=\textwidth]{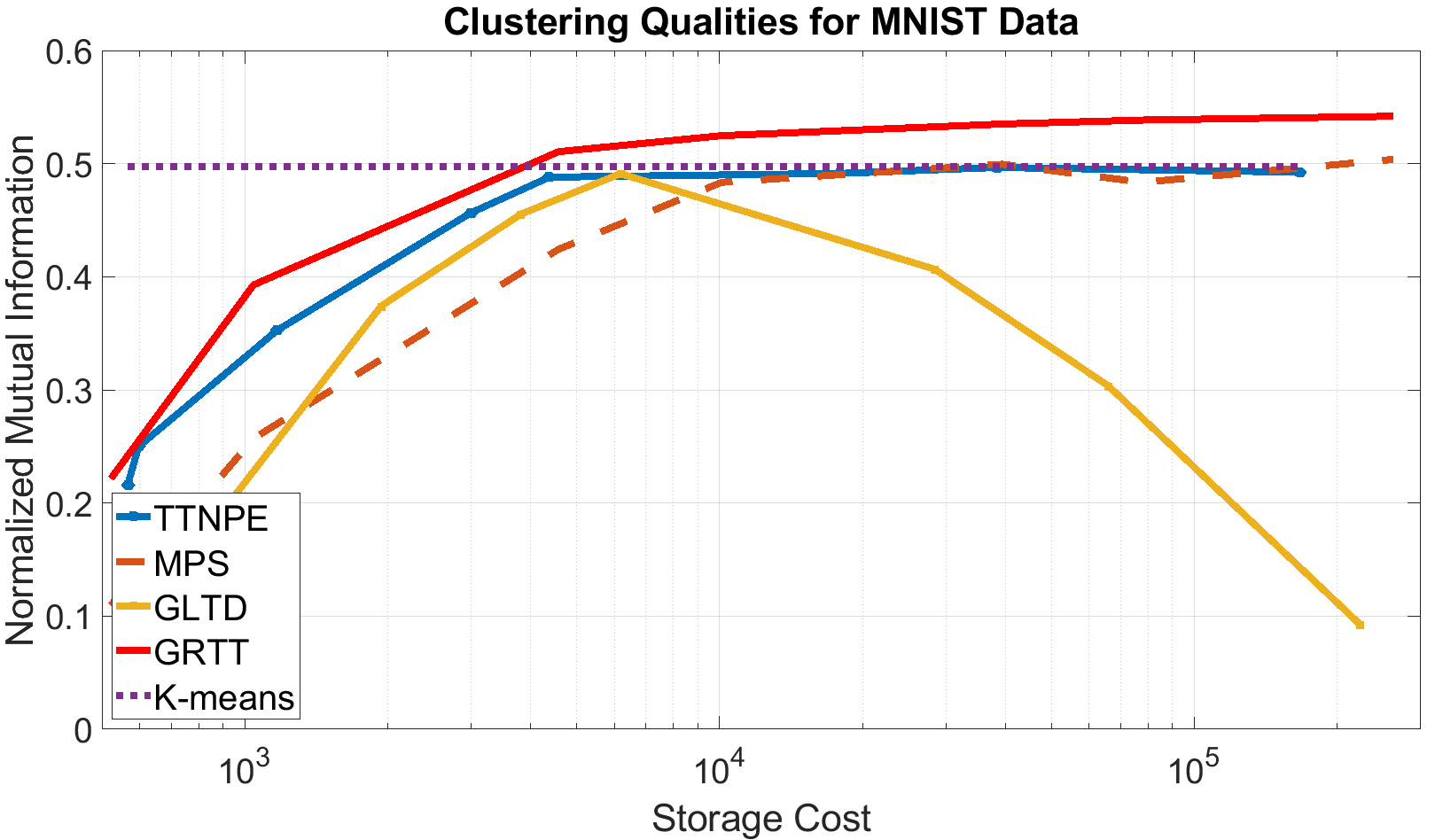}
    \caption{}
    \label{fig:mni_nmi}
\end{subfigure}
\begin{subfigure}[b]{.9\columnwidth}
    \centering
    \includegraphics[width=\textwidth]{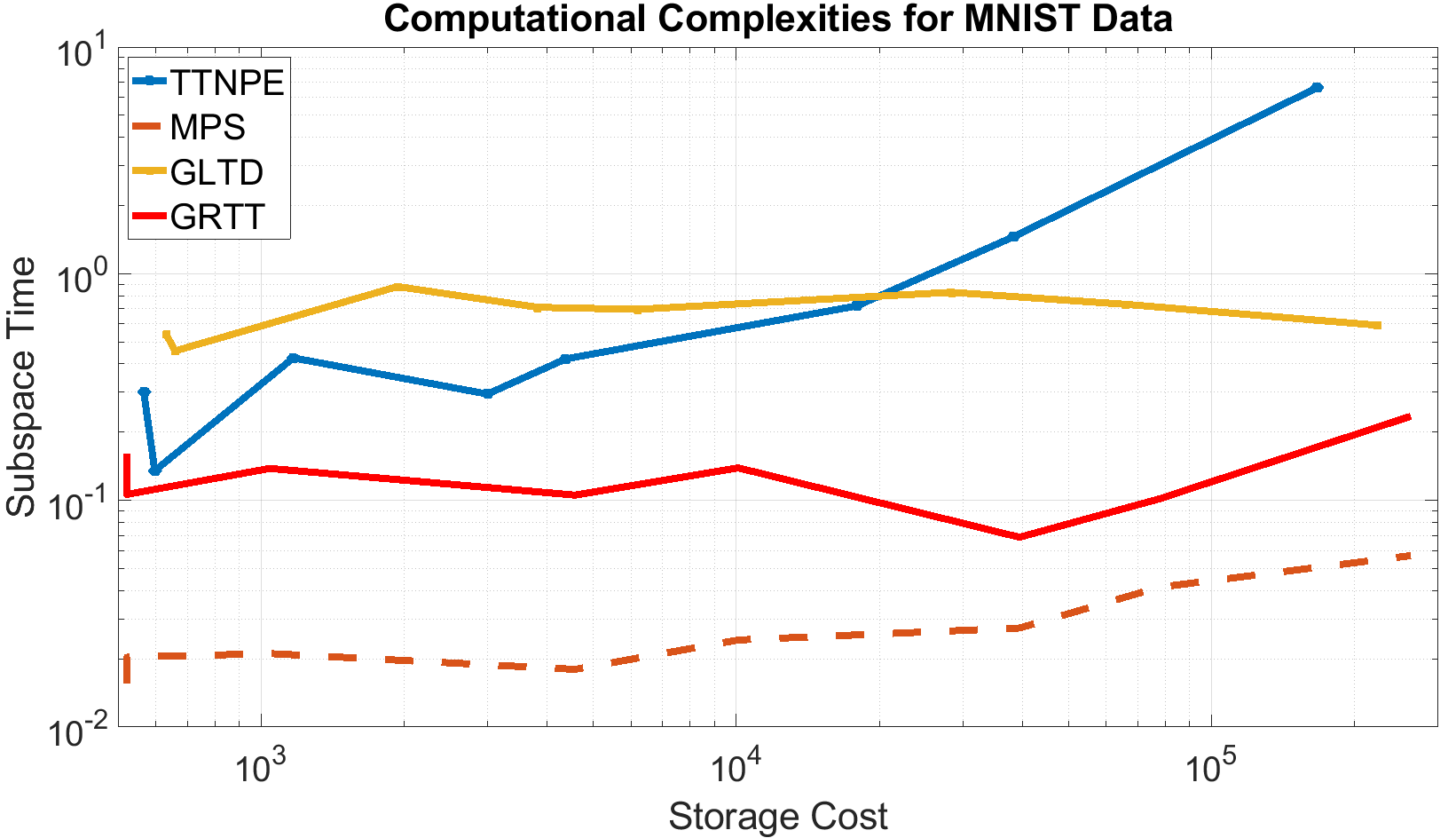}
    \caption{ }
    \label{fig:mni_st}
\end{subfigure}
\caption{(a) Normalized Mutual Information vs. Storage Complexity of different methods for MNIST dataset. (b) Computation Time vs. Storage Complexity of different methods for MNIST dataset.}
\end{figure}
\vspace{-1em}

In Fig. \ref{fig:mni_nmi}, we can see that at all storage complexity levels, our approach gives the best clustering result in terms of NMI. The dotted purple line represents the accuracy of k-means clustering on original tensor data. Even though the performance of TTNPE is the closest to our method, it is computationally inefficient. In Fig. \ref{fig:mni_st}, we can see that our approach is faster than GLTD and TTNPE at all storage complexities. MPS is the most efficient in terms of speed but it provides poor clustering quality.

\subsection{COIL}
The dataset consists of 7,200 RGB images of 100 objects of size $128\times 128$. Each object has 72 images, where each image corresponds to a different pose angle ranging from 0 to 360 degrees with increments of 5 degrees \cite{nenecolumbia}. We used a subset of 20 classes and 32 randomly selected, downsampled, grayscale samples from each class. Each image was converted to an $8\times 8\times 8\times 8$ tensor. 8 samples from each class are used as validation set.

\begin{figure}[htb]
\centering
\begin{subfigure}[b]{.9\columnwidth}
    \centering
    \includegraphics[width=\textwidth]{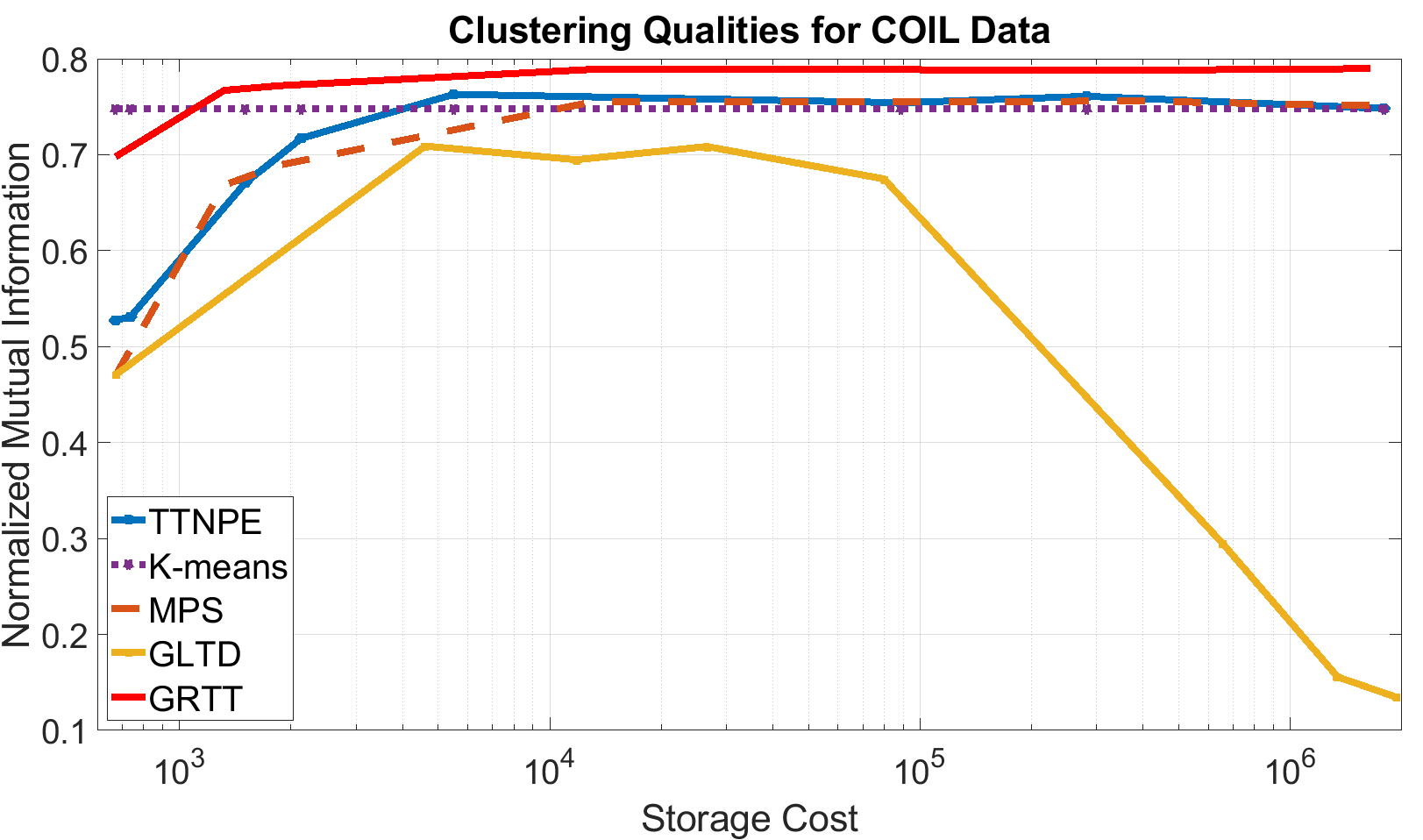}
    \caption{}
    \label{fig:cl_nmi}
\end{subfigure}
\begin{subfigure}[b]{.9\columnwidth}
    \centering
    \includegraphics[width=\textwidth]{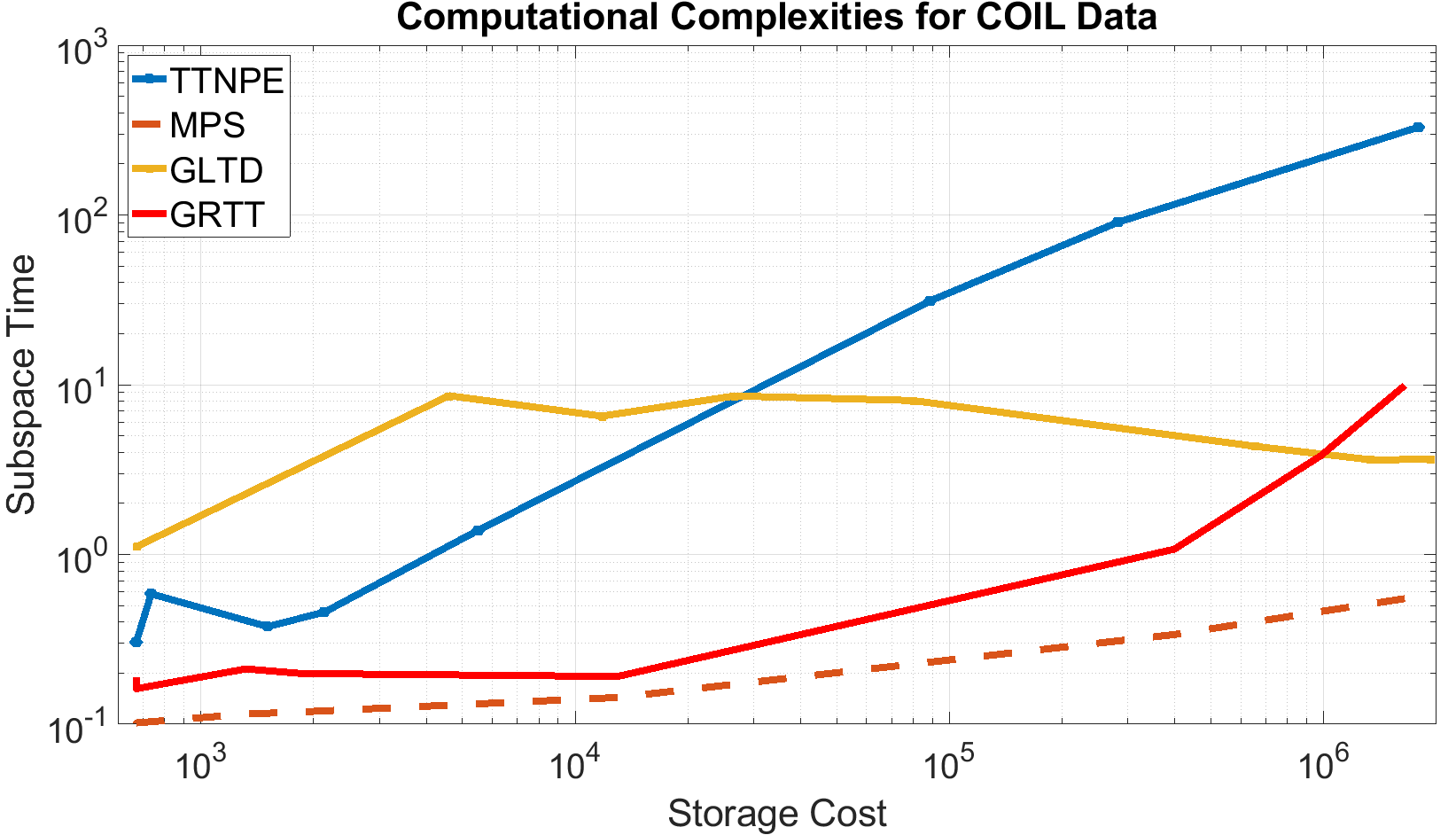}
    \caption{}
    \label{fig:cl_st}
\end{subfigure}
\caption{(a) Normalized Mutual Information vs Storage Complexity of different methods for COIL dataset. (b) Computation Time vs Storage Complexity of different methods for COIL dataset.}
\end{figure}

From Fig. \ref{fig:cl_nmi}, we can see that the proposed method provides the best clustering results compared to all other methods. The results for GLTD seem to deteriorate with increasing ranks, which is a result of using orthonormal tensor factors. TTNPE gives results closest to the proposed method but it is computationally inefficient and gets very slow with increasing $r_n$s. From Fig. \ref{fig:cl_st}, we can see that MPS provides best results in terms of run time but the proposed method has a similar computational complexity while providing better clustering accuracy.

\section{Conclusions}
In this paper, we proposed a unsupervised graph regularized tensor train decomposition for dimensionality reduction. To the best of our knowledge, this is the first tensor train based dimensionality reduction method that incorporates manifold information through graph regularization. The proposed method also utilizes a multi-branch structure to implement tensor train decomposition which increases the computational efficiency. An ADMM based algorithm is proposed to solve the resulting optimization problem. The proposed method was compared to GLTD, TTNPE and MPS for unsupervised learning in two different datasets. 

The proposed method provided the best results in terms of clustering quality while being very efficient in terms of computational cost. The proposed method could also be employed in other dimensionality reduction applications such as denoising, data recovery, and compression. Future work will consider the selection of different design parameters such as the optimal tensor train structure, the construction of the similarity graph and the analysis of convergence rate. Future work will also consider extension of this framework to supervised learning applications.  

\newpage
\bibliographystyle{IEEEtran}
\bibliography{main}
\end{document}